\documentclass{article}
\usepackage{arxiv}
\usepackage{graphicx}
\usepackage{censor}
\usepackage{multirow}
\usepackage{tablefootnote}
\usepackage{amsmath}
\newcommand{\tabref}[1]{Table~\ref{#1}}
\newcommand{\figref}[1]{Figure~\ref{#1}}
\newcommand{\secref}[1]{Section~\ref{#1}}
\newcommand{\subsecref}[1]{Sub-section~\ref{#1}}
\RequirePackage{color}
\definecolor{RED}{rgb}{1,0,0}
\definecolor{BLUE}{rgb}{0,0,1}
\definecolor{White}{rgb}{1,1,1}

\newcommand{\hide}[1]{}

\usepackage{setspace}

\title{Improving Graduation Rate Estimates Using Regularly Updating Multi-Level Absorbing Markov Chains}
\author{
  Shahab Boumi\\ 
  Department of Industrial Engineering\\
  University of Central Florida\\
  Orlando, FL 32826 \\
  \texttt{sh.boumi@knights.ucf.edu} \\
   \And
 Adan Vela \\
  Department of Industrial Engineering\\
  University of Central Florida\\
  Orlando, FL 32826 \\
  \texttt{adan.vela@ucf.edu} \\
}

\begin{document}
\maketitle
\begin{abstract}
American universities use a procedure based on a rolling six-year graduation rate to calculate statistics regarding their students' final educational outcomes (graduating or not graduating). As~an alternative to the six-year graduation rate method, many studies have applied absorbing Markov chains for estimating graduation rates. In both cases, a frequentist approach is used.  For~the standard six-year graduation rate method, the frequentist approach corresponds to counting the number of students who finished their program within six years and dividing by the number of students who entered that year. In the case of absorbing Markov chains, the frequentist approach is used to compute the underlying transition matrix, which is then used to estimate the graduation rate. In this paper, we apply a sensitivity analysis to compare the performance of the standard six-year graduation rate method with that of absorbing Markov chains.  Through the analysis, we highlight significant limitations with regards to the estimation accuracy of both approaches when applied to small sample sizes or cohorts at a university. Additionally, we note that the Absorbing Markov chain method introduces a significant bias, which leads to an underestimation of the true graduation rate.  To~overcome both these challenges, we propose and evaluate the use of a regularly updating multi-level absorbing Markov chain (RUML-AMC) in which the transition matrix is updated year to year.  We empirically demonstrate that the proposed RUML-AMC approach nearly eliminates estimation bias while reducing the estimation variation by more than 40\%, especially for populations with small sample sizes.

\end{abstract}


\section{Introduction}
American universities commonly use a standard six-year graduation rate (SYGR) calculation for reporting their students' outcomes. Based on federal regulations, a~program's graduation rate is defined as the percentage of first-time-in-college (FTIC) students who complete the program within 150\% of the standard enrollment time for the degree \cite{hagedorn2005define}. For~example, for~a four-year program, students~who earn their degrees within six years are considered graduates. The~SYGR method has some disadvantages. Firstly, the~method only considers FTIC students, thereby excluding transfer students, who make up up to 38\% of the student body at many public universities \cite{tugend2018colleges}. In addition, students who complete their programs in more than six years, which is common with students who enroll part-time, are reported as not graduating in this~method.

Based on the definition of the SYGR, an~operational discussion of calculating the SYGR is useful in understanding its features and limitations.  Consider the case of $N^s_y$ FTIC students starting at a university degree program in year $y$.  After~six full years, assume that of the original $N^s_y$ students, $N^g_y$ are observed to graduate.  Accordingly, the~SYGR for the year $y$ is calculated and reported as:
\begin{equation}\label{eq:sygr}
G^r_y = 100 \cdot N^g_y/N^s_y.    
\end{equation}

Immediately, the~first issue with this approach is that the reporting of the graduation rate for a student cohort occurs six years after their initial matriculation in year $y$.  As~such, there is an underlying assumption that students entering the university in year $y+6$ and later will bear out similar results; as~such, the~statistic is arguably  stale.  Moreover, the~accuracy of using the standard SYGR calculation to estimate graduation and retention rates is a direct function of the data available; small data sets produce sensitive estimations. That is to say, estimates of the graduation rate may vary significantly from the {true value} 
The notion of a true value of a graduation rate appears odd in practice, however, here we refer to the true value in the statistical sense as it relates to parameter estimation.

Another common approach for estimating graduation rates is to build a Markov chain based on historical data \cite{nicholls2007assessing}. One advantage of this method over the standard SYGR is that the Markov chain method can be adapted to capture and represent student progress at a university throughout the same six-year period. In~other words, the~method models some temporal aspects of student progress, which SYGR does not model. However, as~we will show in this paper, the~accuracy of estimating graduation rates using Markov chains is quite sensitive to data availability. This disadvantage makes Markov chains unreliable in the context of educational assessments, especially when the sample size of the data used to generate the Markov chains is small.  Additionally, as~part of this paper, we will demonstrate that graduation rates estimated using Markov chains have the potential to be biased, often underestimating the true graduation rate.  As~such, the~driving concern underlying this paper is how small universities should accurately estimate their graduation rates, or~even in the case of larger universities, how they go about estimating their graduation rates for degree programs with lower enrollments (e.g., physics, mathematics) or for cohorts with low representation, e.g.,~women in specific science, technology, engineering, and mathematics (STEM) degree programs \cite{wuhib2014so}.

Consider the case of the   University of Central Florida (UCF)---one of the top five largest universities in America for the last five years  \cite{WinNT}---where only three female students have been observed to both start and graduate from the physics department at UCF between the years of 2008 and 2016.  The~low number is a reflection of multiple factors.  First, the~representation of female students in physics is low; as reported by \cite{porter2019women}, female students only made up 21\% of all physics students across the United States in 2017.  More practically, however, when calculating the SYGR, a~sizable fraction of students are missed because their academic careers will start or end outside the period for which data are available. For~example, over~eight years, the~number of female students reported to declare themselves as physics majors is 79, and~yet for the eight years of available data, the~SYGR can only be calculated for three of the years, corresponding to when students began in 2008, 2009, and~2010.  So even when generously summing and averaging over the three available years, the~reported graduation rate for women in physics would be 18\%, which corresponds to three graduates out of 17 women whose academic records are wholly contained within the eight years of data. Ultimately, the~reliance and reliability of such a metric are questionable, and~more so any implications that might be drawn from~it.

This paper aims to more accurately assess the graduation rate of a university as a whole, as~well as for specific target cohorts. This includes particular majors, under-represented demographic groups, and~transfer students. To~date, prior efforts have dealt with the issues of decreasing data availability according to specifications. In~particular, hierarchical linear models (HLM) have tackled the problem and sought to overcome data availability by understanding particular effects layered on top of main effects \cite{maas2005sufficient,grace2017three,abbott2002influence}.  Like these other methods, our proposed approach uses similar logic to understand how the addition of new information, or~levels of information, can improve graduation rate~estimates.

The remainder of the paper is organized as follows: In \secref{sec:Estimating_graduation_rate}, we show how the accuracy,~in terms of both variance and bias, of~the six-year graduation rate and absorbing Markov chain is a function of data availability.  In~\secref{sec:Methodology}, we explain how our proposed approach reduces variation and bias when estimating graduation rates. Next, the results and a validation analysis are presented in \secref{sec:Results}. Finally, Sections~\ref{sec:Discussion} and \ref{sec:Conclusion} correspond to the discussion and conclusion, respectively.

\section{Estimating Graduation~Rate}  \label{sec:Estimating_graduation_rate}
In this section, we discuss the standard SYGR method and elaborate on the sensitivity of this method.  Similarly, we introduce and compare the usage of an absorbing Markov chain for small cohorts when estimating graduation~rates.

\subsection{Standard Six-Year Graduation~Rate}\label{subsec:6yearsmethod}
Suppose we are interested in estimating a population's six-year graduation rate, $\theta$, given some observed data, $D$. Since only two final six-year outcomes are possible, that is, graduating or not graduating (according to Federal guidelines), each student's outcome can be modeled as a Bernoulli trial. With~this assertion, the~number of students who graduate follows a Binomial distribution with parameter $\theta$. Therefore, the~probability of $k$ students graduating out of $N$, given $\theta$ (the probability of graduating in six years for each student), is:
\begin{equation}\label{eq:binomial}
P(D|\theta) = {N \choose k}\theta^{k}(1-\theta)^{N-k}.
\end{equation}

The standard six-year graduation rate method corresponds to estimating the graduation rate by maximizing the likelihood function in Equation~(\ref{eq:binomial}). Based on the maximum likelihood estimation (MLE), the~estimated graduation rate follows the frequentist approach \cite{heinrich2005parameter}, whereby $\hat{\theta} = k/N$ is an unbiased estimator for graduation~rate. 

\begin{figure}[h!]
\centering
\includegraphics[width = 0.75\textwidth]{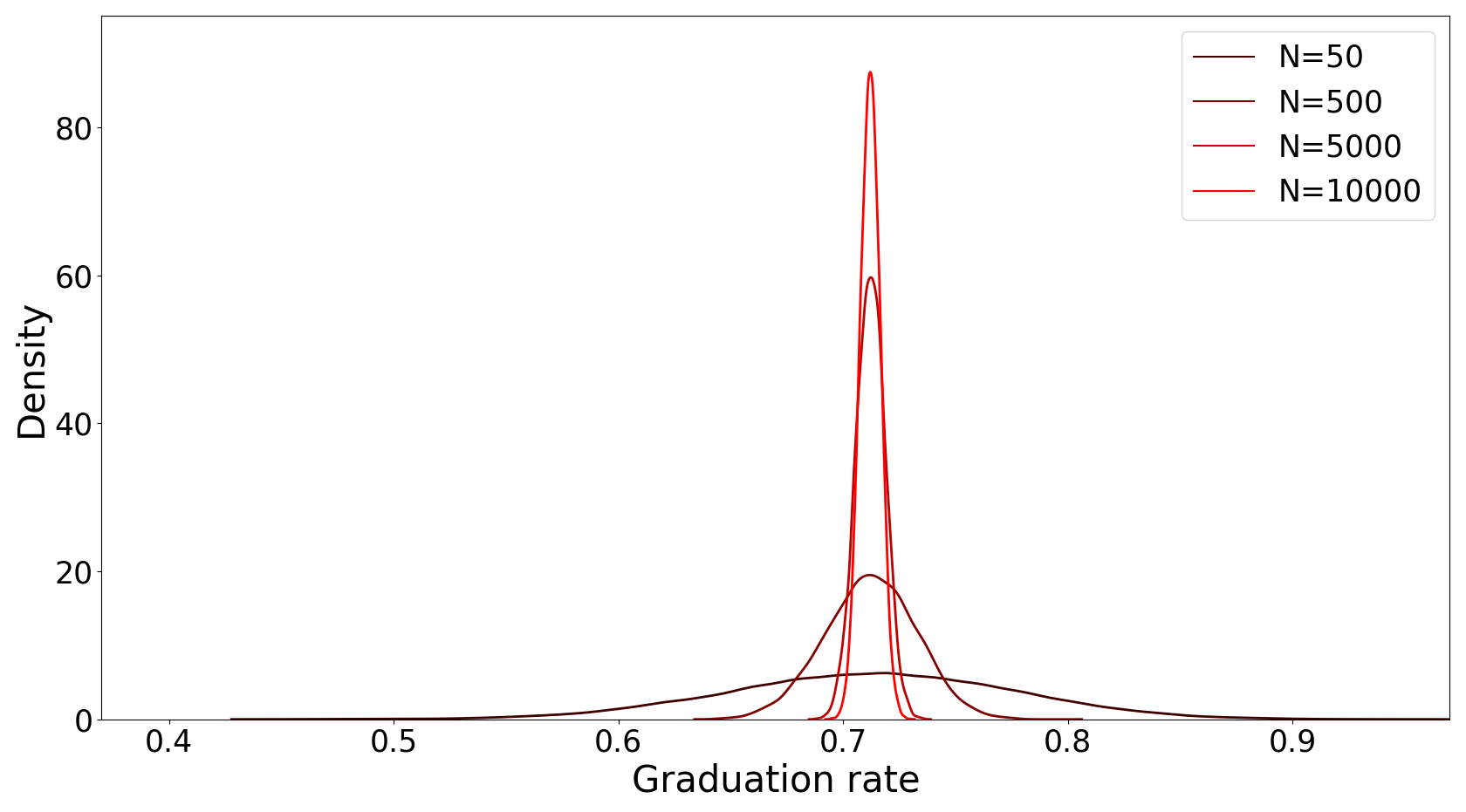}\hspace{5pt}

\caption{The 
 probability density function (using Gaussian kernel smoothing with $\sigma$  =  1) for estimations using the six-year graduation rate (SYGR)~method.} \label{fig:six_years_Hist}
\end{figure}

In order to demonstrate the performance of the MLE approach, we use data collected from the University of Central Florida (UCF) 
Based on historical data, the~six-year graduation rate at UCF for first-time-in-college (FTIC) students starting in 2008 is 71.2\%, with 28.8\% of students graduating in more than six years or halting enrollment. Assuming 71.2\% to be the true value, parameterizing a binomial distribution representing the number of students graduating within six years (i.e., $\theta = 0.71$), we randomly simulate 10,000 student cohorts of different sample sizes, $N$, with~the resulting SYGR calculated using Equation~(\ref{eq:sygr}). Here, each one of the 10,000 cohort samples represents an incoming Freshman class, or~perhaps a specific cohort (e.g., women in STEM fields). The~simulated numbers of graduates in the cohorts with different sample sizes alongside the corresponding average graduation rate and standard variation across the cohort samples are summarized in \tabref{tab:sixyearsgratesamples}. The~corresponding probability density function (pdf) of the six-year graduation rate for each set of cohorts with a fixed size, representing an estimate, is shown in \figref{fig:six_years_Hist}. As~indicated in \figref{fig:six_years_Hist} and   \tabref{tab:sixyearsgratesamples}, the~distribution of graduation rate estimates for the cohorts with small sample sizes can vary significantly from the asserted true value of 71.20\% (see the case for $N = 50$ where $s = 6.36\%$). That said, the~sample mean of the graduation rates over the 10,000 simulated cohorts is between 71.19\% and 71.21\%, which~closely matches the actually reported graduation rate of 72.20\%, again indicating that the maximum likelihood estimator is an unbiased estimator. Both the figure and table illustrate how the sample standard deviation for cohorts with small sample sizes ($N = 50$) is significantly larger when compared with larger sample sizes ($N = 5000$), i.e.,~6.4\% versus 0.6\%. Accordingly, when $N = 50$ and $N = 500$, the~probability that the SYGR is reported to be lower than 66\% or higher than 76\% is 0.53 and 0.03, respectively (corresponding to a graduation rate reporting error of greater than 5\%).  These differences can be considered significant in the context of college rankings and when being evaluated by government or accreditation boards. And while the numerical values above are specific to UCF, similar sensitivity results are expected at other American universities. 
Accordingly, it is arguably true that natural statistical variations have the potential for an oversized impact on the ranking or perception of small departments or small colleges when compared to larger departments or~colleges.

\unskip

\begin{table}[h!]
\caption{The number of graduates, average graduation rate, and~standard deviation of estimated graduation rate for cohorts with different sample~sizes.}
\label{tab:sixyearsgratesamples}
\centering
\begin{tabular}{|c|c|c|c|c|}
\hline
 \textbf{Sample Number} & \textbf{N = 50} & \textbf{N = 500} & \textbf{N = 5000} & \textbf{N = 10,000}  \\
\hline
1 & 40 & 364 & 3572 & 7085 \\ \hline
2 & 36 & 354 & 3578  &7062  \\ \hline
3 & 35 & 371 & 3544 & 7103  \\ \hline
4 & 32 & 363 & 3525 & 7186  \\ \hline
5 & 36 & 383 & 3571  &7092  \\ \hline
. & . & . & .  &.  \\ \hline
. & . & . & .  &.  \\ \hline
 10,000 & 38 & 364 & 3542 &  7105  \\\hline

Sample average Grad. Rate &71.19\% & 71.21\% & 71.20\% & 71.20\% \\ \hline
Sample standard deviation & 6.36\% & 2.04\% & 0.64\% & 0.45\%  \\ 
\hline
\end{tabular}
\end{table}
\unskip

\subsection{Absorbing Markov~Chain}\label{subsec:AMC}
Different approaches are used to evaluate students' performance and persistence in higher education systems, among~which machine learning algorithms and stochastic models are the most common~\cite{asee_peer_34921,aghajari2020decomposition,saa2016educational,mirzaei2020detecting,mirzaei2019modeling,ebrahiminejad2019pathways,zahedi2020multi}. Markov models have been used in many educational studies to analyze students' academic progress and behaviors \cite{nicholls2007assessing,boumi2020quantifying,nicholls2009use,shah1999undergraduate,al2007application,boumi2019application}. For~example, \cite{nicholls2007assessing} analyzed the progress of graduate students  through a degree program in Australia.~Using Markov modeling techniques, the~authors assessed students' performance according to measures like expected time to graduation and graduation rate. The~authors developed a Markov model that includes two absorbing states representing withdrawing from the program and graduating.  Additional transient states represent the students' status at the end of each year based on their academic performance.  A~similar modeling procedure is provided by \cite{bairagi2017stochastic}, who focused on a university system in northern India.  While maintaining a significantly different academic structure (e.g., number of courses per semester, semester exams for each course), their Markov models were also able to track and model student~progression. 

\begin{figure}[h!]
\centering
\includegraphics[width = 0.70\textwidth]{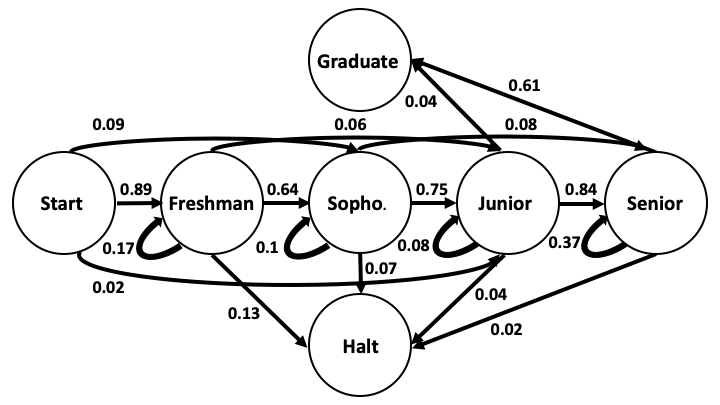}\hspace{5pt}

\caption{Representation of Markov chain transitions at~the University of Central Florida (UCF).} \label{fig:Transition_flow}
\end{figure}

For the prior works cited above, the authors made use of a specific class of Markov chains referred to as absorbing Markov chains (AMCs).  Absorbing Markov chains have two classes of states: transient states and absorbing states. In~the case of applying AMCs to track student progress through a degree program, the~total number of states (both absorbing and transient) for an AMC is typically finite.  For~modeling American four-year universities with an AMC, absorbing states could correspond to graduating or halting. In~contrast, transient states could correspond to academic level (e.g., Freshman, Sophomore, Junior, Senior).  An~example application of an AMC for an American four-year university is provided in \figref{fig:Transition_flow}.  In~the case of an AMC, when the system transitions from a transient state to one of the absorbing states, it cannot exit the state.  Again, transitioning to an absorbing state corresponds to a student halting their education or graduating; however, practically speaking, a~student could always earn another degree or re-enroll years later.  In~addition to a list of absorbing and transition states, each~AMC, like~any other Markov model, is defined by a transition matrix $P_{ij}$ representing the probability of moving from state $i$ to state $j$ \cite{shah1999undergraduate,boujelbenedata}. The~canonical form of an absorbing Markov chain with $r$ absorbing states and $t$ transient states is based on the matrix $P$ in (\ref{eq:Transition_Mat}). In~the matrix, $R$ is a $t\times r$ matrix  containing the transition probabilities from the transient states to the absorbing states, $Q$ is a $t\times t$ matrix that represents transition probabilities within the transient states, $I$ is an $r\times r$ identity matrix, and~$0$ is an $r\times t$ zero matrix that allows for the AMC to model the trapped dynamics when entering an absorbing state \cite{brezavvsvcek2017markov,ledwith2019ethics}. The matrix $P$ is used as part of the dynamic equation $x(y+1)=x(y+1)P$, where $x(y)$ is a vector containing the probability distribution over the states at time-step $y$.  In our case, the dynamic equation probabilistically describes how a student advances through their academic career.  While matrix $P$ provides the one-step transition probabilities between states, the~matrix power $P^n$ represents $n$-step probabilities of transitions between states. In~other words, $[P^{n}]_{i,j}$~is the probability that a system that is initially in state $i$ will be in state $j$ after exactly $n$ \mbox{steps \cite{ross2014introduction,polyzou2019scholars,hadad2020source}}. Each~absorbing Markov chain has two useful calculated properties: expected time until absorption (U) and probabilities of absorption (B) to the absorbing states~\cite{eledum2019undergraduate}. These characteristics are computed with Equations~(\ref{eq:U_Mat}) and (\ref{eq:B_Mat}).  In~our case, their values correspond to the probability and expected time to graduate or to halt.
\begin{equation}P = \begin{bmatrix} \label{eq:Transition_Mat}
Q & R \\
0 & I \\
\end{bmatrix}\end{equation}
\begin{equation} \label{eq:U_Mat}
    U = N\times 1
\end{equation}
\begin{equation} \label{eq:B_Mat}
    B = N\times R
\end{equation}
where \begin{equation*} \label{eq:Fundemental_Mat}
    N = (I-Q)^{-1}
 \end{equation*}

In order to demonstrate how absorbing Markov chains are used to estimate graduation rates, we consider students' academic levels (Freshmen, Sophomore, Junior, Senior) as the transient states and students' final educational outcomes (graduate or halt) as the absorbing states. All students start from a dummy state (the start
state); then, based on their incoming academic credits (e.g., Advanced Placement credits), they are assigned to other states. After~this initial assignment, students then advance through the transient states based on their accumulation and successful completion of academic credits.  Ultimately, students are absorbed into one of the absorbing states.  For~our purposes, when processing historical data, we declare students to have halted their education if they do not enroll for three consecutive semesters. The~possible transitions and transition probabilities for students who started their education in Fall 2008 at UCF are shown in \figref{fig:Transition_flow}.  As~typically done, the~transition probabilities between states are extracted from historical data using the maximum likelihood estimator, resulting in an unbiased estimation of individual transition probabilities. Each state in the AMC corresponds to the student's academic standing at the end of each academic year. For~example, at~the end of one year, 10\% of sophomore students remain at the sophomore level, while 75\% and 8\% of the students advance to a junior and senior academic standing, and~finally, 7\% will halt their education. In~order to find the fraction of students who  graduated within six years, we need to calculate $P^{6+1}$ ($+1$ accounts for students beginning in the {start} state) and observe the entry that contains the transition probability from  the {start} state to the graduate state.  Assuming that the AMC and its transition probabilities (illustrated in \figref{fig:Transition_flow}) are an accurate representation, the~$N$-step transition matrix is given by the matrix in Equation~(\ref{eq:Transition_Mat_power7}) below, with~the bolded value $0.686$ corresponding to the estimated graduation rate.

\begin{equation}P^7 = \begin{bmatrix} \label{eq:Transition_Mat_power7} 
0 ~~& 0 ~~& 0 ~~& 0.002 ~~& 0.059    ~~& 0.253    ~~& \textbf{0.686} \\
0 ~~& 0 ~~& 0 ~~& 0    ~~& 0.025    ~~& 0.270 ~~& 0.705 \\
0 ~~& 0    ~~& 0 ~~& 0 ~~& 0.008    ~~& 0.144 ~~& 0.848 \\
0 ~~& 0    ~~& 0    ~~& 0 ~~& 0.003 ~~& 0.073 ~~& 0.924 \\
0 ~~& 0    ~~& 0    ~~& 0    ~~& 0 ~~& 0.031 ~~& 0.969 \\
0 ~~& 0    ~~& 0    ~~& 0    ~~& 0    ~~& 1    ~~& 0 \\
0 ~~& 0    ~~& 0    ~~& 0    ~~& 0    ~~& 0    ~~& 1 \\
\end{bmatrix}\end{equation}

In order to evaluate the performance of the AMC method (i.e., using $P^{6+1}$) to estimate graduation rates, 10,000 cohorts with different sample sizes ($N$ = 50, 500, 5000, 10,000) are generated based on the UCF transition matrix parameters depicted in \figref{fig:Transition_flow}; again, this assumes that the values of transition matrix are the true values.  The~academic trajectory of the students is sampled directly from the Markov model. Examples of sampled generated students' academic trajectories are provided in \tabref{tab:trajectory_example}.  For~each of the 10,000 sets of $N$ generated student trajectories, a~unique transition matrix, $\hat{P}$, is~generated based on the sampled generated data.  So, for $N = 50$, there are 10,000 induced transition matrices $\hat{P}$ whereby the transition probabilities are calculated based on academic trajectories using only 50 students.  The~estimated graduation rate for a sample of 50 students is then extracted from the estimated $\hat{P}^{6+1}_{start,graduate}$; this process is repeated 10,000 times, with~a report on the sample  in \tabref{tab:AMCsamples}.

\begin{table}[h!]
\caption{Example of year-to-year progressions of~students.}
\label{tab:trajectory_example}
\centering
\begin{tabular}{|c|c|}
\hline
\textbf{Student Number}	& \textbf{Year-to-Year Progression}	\\
\hline
1 & Fr-So-Ju-Sn-Sn-Gr\\\hline
2 & Fr-Fr-So-So-So-So-Ht\\\hline
3 & Fr-Fr-Ht\\

\hline

\end{tabular}
\end{table}

\begin{figure}[h!]
\centering
\includegraphics[width = 0.75\textwidth]{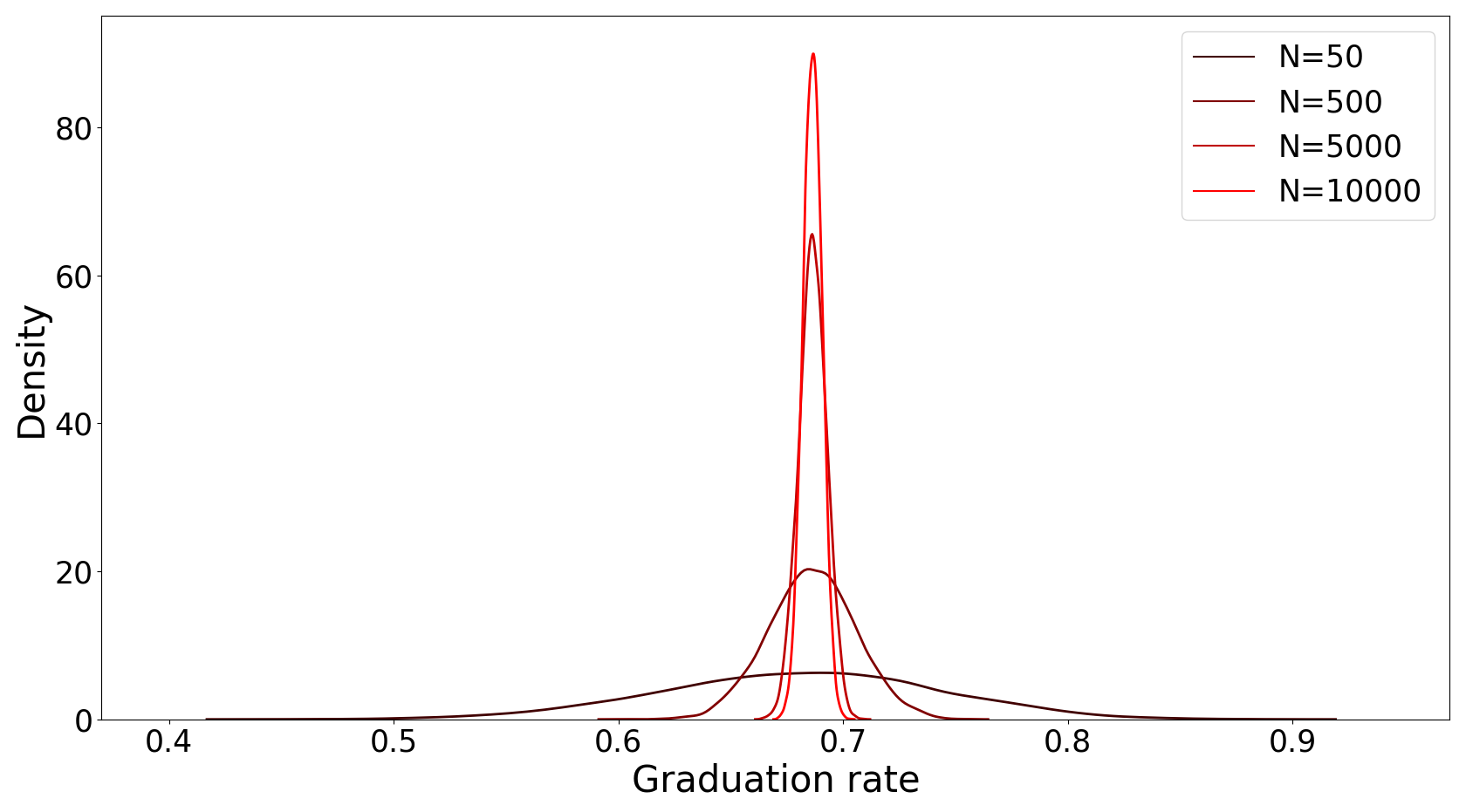}\hspace{5pt}

\caption{The empirical probability density function of the estimated graduation rate using the absorbing Markov chain~method.} \label{fig:MC_density}
\end{figure}

The sampled probability distribution functions (pdfs) of the estimated graduation rates when using the AMC method are shown in \figref{fig:MC_density} for a variety of cohorts with different sample sizes. As~the figure illustrates and \tabref{tab:AMCsamples} reports, the~sample standard deviation of the estimated graduation rate for cohorts with small sample sizes is high (e.g., for~$N = 50$, the~sample standard deviation is $s = 6.30\%$). Besides~the poor performance of the AMC in terms of limiting the sample standard deviation, the AMC~introduces a bias from the true graduation rate, as established by the actual six-year graduation rate. The~estimated graduation rate based on the AMC method is 68.6\% (the bold number in Equation~(\ref{eq:Transition_Mat_power7})), while the true six-year graduation rate for the same cohorts on which the original was based  is 71.20\%; the same bias is present for all sample sizes. In~other words, the~AMC model underestimates the six-year graduation~rate.  

\unskip

\begin{table}[h!]
\caption{$\hat{P}^{6+1}_{start,graduate}$.}
\label{tab:AMCsamples}
\centering
\begin{tabular}{|c|c|c|c|c|}
\hline
 \textbf{Sample set} & \textbf{N = 50} & \textbf{N = 500} & \textbf{N = 5000} & \textbf{N = 10,000}  \\
\hline
1 & 74.17\% & 70.13\% & 69.23\% & 68.70\% \\ \hline
2 & 59.17\% & 65.08\% & 69.38\%  &68.99\%  \\ \hline
3 & 69.32\% & 69.11\% & 68.47\% & 68.14\%  \\ \hline
4 & 66.62\% & 67.40\% & 69.03\% & 67.46\%  \\ \hline
5 & 78.78\% & 70.68\% & 68.25\%  &69.25\%  \\ \hline
. & . & . & .  &.  \\ \hline
. & . & . & .  &.  \\ \hline
 10,000 & 75.30\% & 65.58\% & 68.75\% &  68.12\%  \\\hline

Sample average Grad. Rate & 68.64\% & 68.65\% & 68.64\% &  68.65\% \\ \hline
Sample standard deviation & 6.30\% & 1.92\% & 0.62\% &  0.43\% \\ 
\hline

\end{tabular}
\end{table}

\begin{figure}[h!]
\centering
\includegraphics[width = 0.7\textwidth]{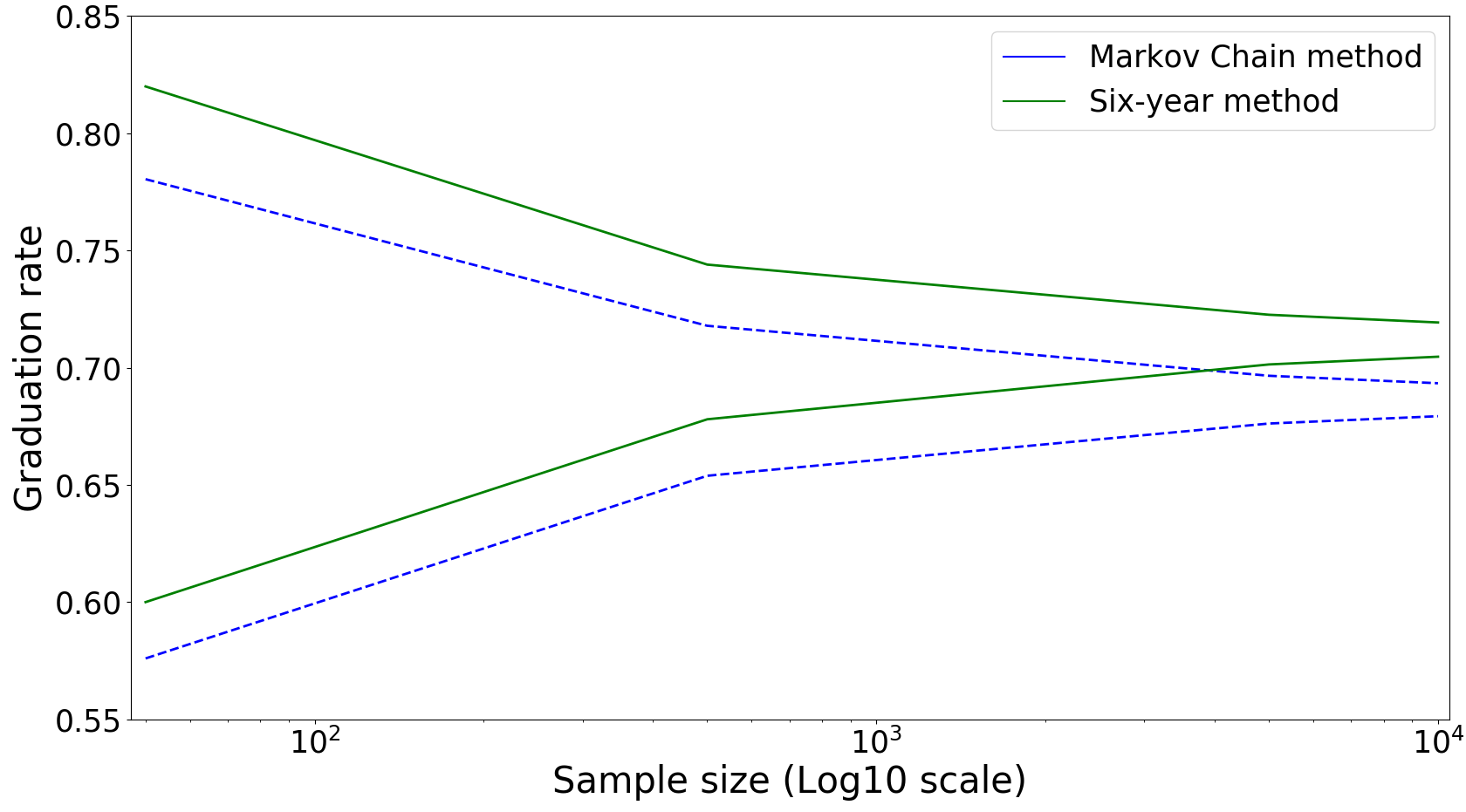}\hspace{5pt}
\caption{The 5--95\% inter-quartile range for graduation rates of cohorts with different sizes obtained by the absorbing Markov chain and six-year graduation rate~methods.} \label{fig:SixYear_MC_Percentiles}
\end{figure}

\figref{fig:SixYear_MC_Percentiles} compares 5--95\% inter-quartile interval as a measure of performance (in terms of estimation variation and bias) for both the absorbing Markov chain and six-year graduation rate methods. Based on the results, we see that the sample standard deviation of the estimated graduation rates in both the SYGR and AMC methods is higher for cohorts with small sample sizes.  Furthermore, the AMC has a 2.6\% (71.2--68.6\%) bias in estimating the true graduation rate, unlike the SYGR method, which does not have an estimation~bias.

To overcome the challenges presented above (i.e., bias and large sample standard deviation for small sample sizes), we propose the use of a regularly updating multi-level absorbing Markov chain (RUML-AMC) method to cope with these shortcomings (in terms of both variation and bias).  We then provide a sensitivity analysis to demonstrate the benefit of this method in improving the accuracy of graduation rate estimates. The~details of the proposed methodology are explained in the next~section.

\section{Methodology}\label{sec:Methodology}
In this section, we propose two techniques to overcome shortcomings related to high sample standard deviation and bias when estimating graduation rates using AMCs. The first technique overcomes estimation bias by expanding an AMC model to include multiple levels for each academic standing (e.g., Freshman, Sophomore). The second technique, focusing on reducing the sample standard deviation of the estimated graduation rate, is based on regularly updating the transition matrix as new data becomes available, even if the data is incomplete. In combination, these contributions help us estimate graduation rates with lower bias and smaller sample standard deviation than the more traditional SYGR and AMC methods discussed in the previous section.

\subsection{Reducing Estimation Bias}\label{subsec:Bias}
In \secref{sec:Estimating_graduation_rate}, we indicated that there is a noticeable difference in the expectation of the estimated graduation rate when using absorbing Markov Chains (68.6\%) as compared to the six-year graduation rate method (71.2\%). This bias is caused by the underline assumption in the absorbing Markov model that a student will remain at the same academic level (i.e., state) year-on-year with the same probability; this phenomenon is an expression of the Markov property). This assumption is unrealistic as the probability of halting enrollment or advancing academic levels changes as students spend additional years at the same academic level. As an example, the transition probability for students moving from Freshman state to Sophomore depends on how long the student has been classified as a Freshman.  So for example,
\begin{equation}\label{eq:Markovian_prob}
P(\textrm{Fr. to So. $|$ 1 year in Fr.})\\
\neq
\\
P(\textrm{Fr. to So. $|$ 3 years in Fr.})\text{,}
\end{equation}
which effectively states that the probability of a student advancing from the Freshman level to the Sophomore level depends on how many years they have been categorized as a Freshman.  The approximation of non-Markovian behavior of student advancement through academic levels ultimately leads to the Markov model incorrectly estimating the true value of the graduation rate.

\begin{figure}[h!]
\centering
\includegraphics[width=0.70\textwidth]{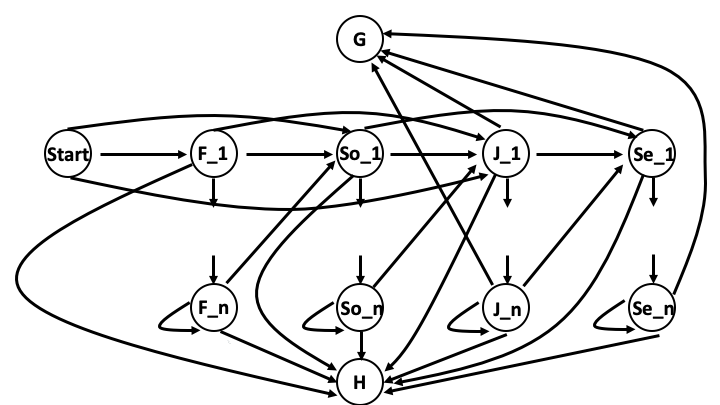}\hspace{5pt}
\caption{Representation of a multi-level absorbing Markov chain transitions with $n$ numbers of sub-states.} \label{fig:Multi_level_transition}
\end{figure}

To tackle this issue, we propose the use of a multi-level absorbing Markov chain (ML-AMC) with the addition of sub-states for each transient state corresponding to academic levels; each sub-state will corresponding to the number of years a student has spent at a particular academic level. The general form of the transition flows for this absorbing Markov chain is illustrated in \figref{fig:Multi_level_transition}. In the figure, we have defined $n$ levels for each transient state in which only the last sub-state ($n$th) of each academic level has a self-loop transition.
For example, in the case $n$=2, the academic trajectory for student number 2 in \tabref{tab:trajectory example} is: $Fr_1$-$Fr_2$-$So_1$-$So_2$-$So_2$-$So_2$-$Ht$.

In augmenting the number of states associated with each academic level, we are able to account for the discrepancy noted in Equation \ref{eq:Markovian_prob}.  Using historical data, we calculate the transition probabilities between different academic levels, given the number of years a student stays in the same state before the transition to the next (Table \ref{tab:TransitionMatrixN3}). As we see in the table, students' academic level advancement does not follow a Markov chain behavior. For example, the transition probability from Freshman to Sophomore given staying in the Freshman state for one year is 64\%, while the same transition probability for students who stay in Freshman state for three years decreases to 40\%.

\begin{table}
\caption{Transition matrix for the absorbing Markov chain with 3 years remaining in a given states}
\centering
\begin{tabular}{|c|c|c|c|c|c|c|}\hline
{}&\textbf{$F_1$}	& \textbf{$F_2$} & \textbf{$F_3$} & \textbf{$So_1$} & \textbf{$So_2$} & \textbf{$So_3$}	\\
\hline
$F_1$  & 0\%  & \textbf{21}\% &0\%& \textbf{64}\% &0\% &0\%\\ \hline
$F_2$ & 0\%&0\%& \textbf{1}\%& \textbf{63}\%&0\%&0\%\\ \hline
$F_3$ & 0\%&0\%&0\%&\textbf{40}\%&0\%&0\% \\ \hline
\end{tabular}
\label{tab:TransitionMatrixN3}
\end{table}

\begin{figure}[h!]
\centering
\includegraphics[width=0.75\textwidth]{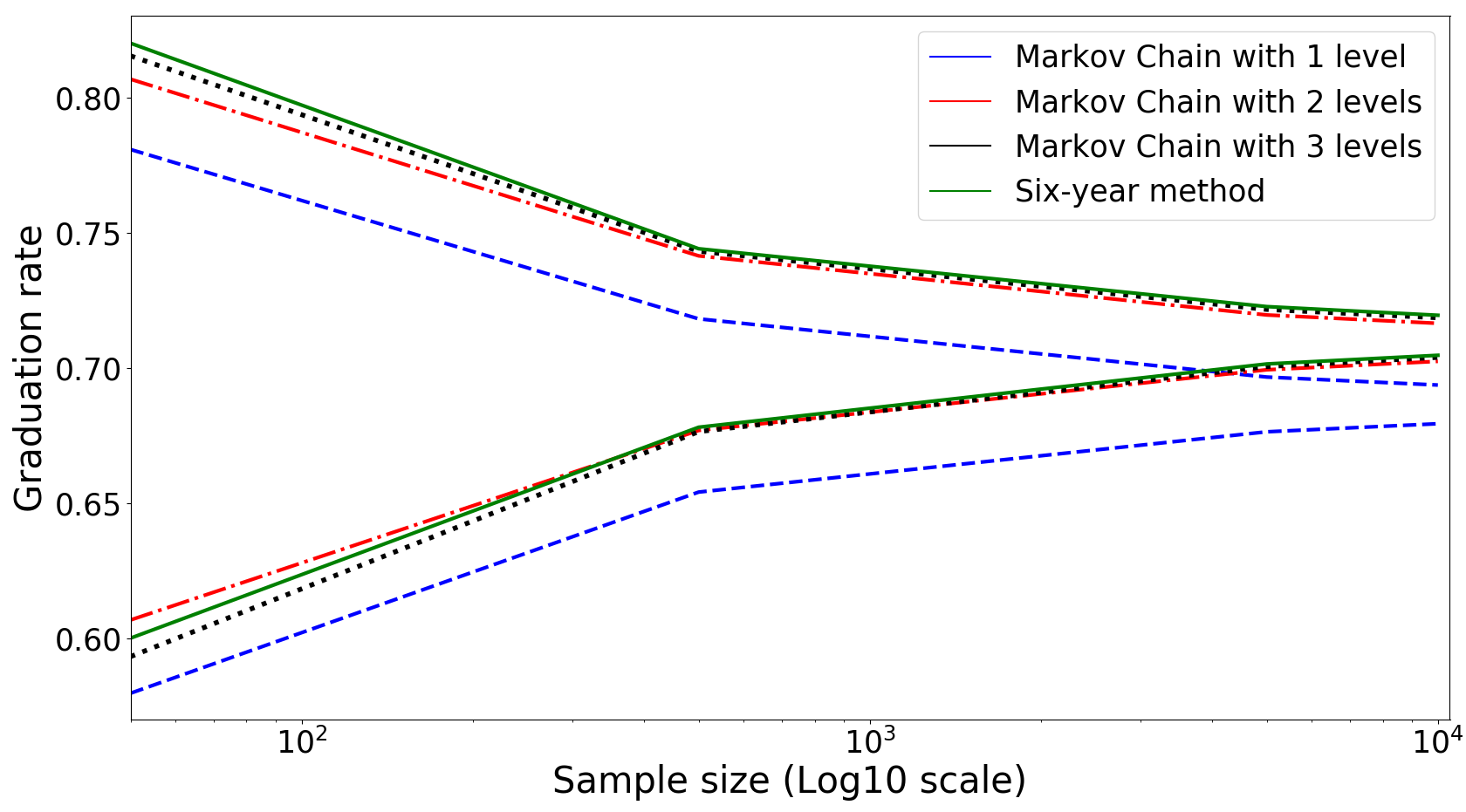}\hspace{5pt}
\caption{5\%-95\% inter-quartile range for graduation rate of cohorts with different size obtained by SYGR, AMC, and ML-AMC with $n=2$ and and $n=3$} \label{fig:All_methods_percentiles}
\end{figure}

The results for graduation rate estimation using SYGR, AMC, and the multi-level absorbing Markov chain with $n$=2 and $n=3$ are shown in \figref{fig:All_methods_percentiles}. As shown empirically in the figure, the estimation bias for the Markov chain method with $n=2$ and $n=3$ levels becomes negligible compared to the Markov chain with $n=1$ level, which corresponds to the standard AMC. While the estimation bias is virtually removed through the addition of multiple levels, the sample standard deviation of the graduation rate estimates still remains high, especially for small $n$, as represented by the size of the 5\%-95\% inter-quartile spread.  In the next sub-section, we address this shortcoming when applying Markov chains to estimate graduation rates.

\subsection{Reducing the Sample Standard Deviation in Estimates}\label{subsec:Variance}
In the previous sub-section, we proposed a multi-level absorbing Markov chain (ML-AMC) approach to reduce estimation bias. In this sub-section, we apply a regularly updating multi-level absorbing Markov chain (RUML-AMC) approach to update the transition probabilities with the addition of data on a year-by-year basis over a six period. In this approach, we assume the data to calculate the transition probabilities for all states is null, and as new students join the degree program at a constant rate, which is equal to the initial number of students enrolled, then the transition probabilities are updated for each state. For example, if 50 students initially enroll in a degree program, the total number of enrolled students at the beginning of the second year is assumed to be $50+ 50 \times$ <Freshman retention rate>, which includes new students and the student that remain in college.

In this method, given the additional observations of new students during a fixed six-years horizon, the transition probabilities between every two consecutive states are learned and updated year-by-year. That implies more learning happens at earlier states (e.g., Freshman and Sophomore) where the model receives more observations. \tabref{tab:regularly_MC_method_sample_size} illustrates an example of sample sizes (number of students) observed for each state in different years for a RUML-AMC with $n$=1. As the table indicates, the size of the data used to calculate the transition probabilities for the Freshman and Sophomore states is larger when compared to the Junior and Senior states from the first year to the sixth year.  The increase in student samples for these earlier states helps to reduce overall model uncertainty, which ultimately reduces the sample standard deviation of the estimates of the graduation rate.  In fact, the reduction in uncertainty is driven by a significant increase in data for the Freshman and Sophomore years (270 and 194 observations within 6 years) to reduce the uncertainty in the transition probabilities.  The Freshman and Sophomore states introduce the greatest variance as the advancement rates are 0.64 and 0.75 according to the standard AMC, see \figref{fig:Transition_flow}; as the true transition probabilities approach 0.5, the standard deviation in estimating the parameter increases. The best way to manage this associated increase in sample standard deviation ($s=\sqrt{p(1-p)/N}$) is to increase the number of samples, $N$, which is accomplished by including new data as it appears each year.
%
%

\begin{table}
\caption{Number of students observed in each state for different years with RUML-AMC and $n$=1}
\centering
\begin{tabular}{|c|c|c|c|c|c|c|}\hline
{}&Year 1	& Year 2	& Year 3	& Year 4	& Year 5	& Year 6	\\
\hline
$Fr$  & 47  & 93 & 140 & 186 & 232 & 270 \\ \hline
$So$ & 40 &  80 & 120 & 158 & 190 & 194 \\  \hline
$Ju$ & 36 &  72 & 106 & 134 & 140 & 141 \\ \hline
$Se$ & 19 &  36 & 48 & 51 & 52 & 52 \\ \hline
$H$ & 13 &  25 & 38 & 51 & 94 & 95 \\ 
\hline
\end{tabular}
\label{tab:regularly_MC_method_sample_size}
\end{table}

Results for the sample standard deviation of the estimated graduation rate when applying this technique over a fixed six-year period are provided in \tabref{tab:std_comparison_the_MLAMC_model_n1}. As we see in the table, our first estimation has a large standard deviation as a direct result of the small number of students that are used to estimate the transition probabilities in the Markov chain; this is equivalent to the standard AMC discussed in \subsecref{subsec:AMCs}. For the second estimation, given that 50 new students are added to the previous pool of students, the sample standard deviation of the transition probability estimates is reduced, and as such, the corresponding pdf for the sampled estimated graduation rate is narrowed compared to the first estimation.  This phenomenon of shrinking sample standard deviation of the estimated graduation rate is repeated with the introduction of new student data each year. Finally, the sixth estimation, which uses the transition matrices of five previous years, provides the most accurate measure. As we observe in the table, the sampled standard deviation for the six-year graduation rate method (6.4\%) is reduced by more than 40\% compared to sixth-year estimation using the regularly updating absorbing Markov chain method (3.4\%).   

The use of this technique comes at no particular time-cost as all the sampled data remains within the same 6-year time period.  For the cohort entering in year $y$, there is 6 years of data, while for the cohort entering in year $y+1$, there is 5 years of data.  As such, a reduction in the sampled standard deviation of the estimated graduation rate does not require collecting data over additional years.  This is in contrast to performing a $n$-year rolling average of SYGR rate given by $\hat{G}^r_{y-y+n}=100*\frac{\sum_i=0^{n-1}{N^G_y}}{\sum_i=0^{n-1}{N^s_y}}$.  For each additional piece of data that is average, computation of the statistic requires a delay of 1 year, and even then, the benefit is limited.  So, for example, if the SYGR is averaged over two and three years (based on the 10000 simulated cohorts of $N=50$), the sample standard deviation of the estimates is 4.6\% and 3.7\%, as compared to 4.5\% and 3.9\% for the RUML-AMC when using two and three cohorts of students.

\begin{table}
\caption{Standard deviation of estimated graduation rate from year 1 to 6 for cohort with $N$=50 for the RUML-AMC with $n$=1}
\centering
\begin{tabular}{|c|c|c|} \hline
 Estimation number & Sampled Std. of Estimates	\\ 
\hline
1 & 6.2\%\\ \hline
2 & 4.5\% \\ \hline
3 & 3.9\%\\ \hline
4 & 3.6\%\\ \hline
5 & 3.5\%\\ \hline
6 & 3.4\%\\ 
\hline
\end{tabular}
\label{tab:std_comparison_the_MLAMC_model_n1}
\end{table}

\section{Results and~Validation} \label{sec:Results}
\begin{figure}[h!]
\centering
\includegraphics[width = 0.75\textwidth]{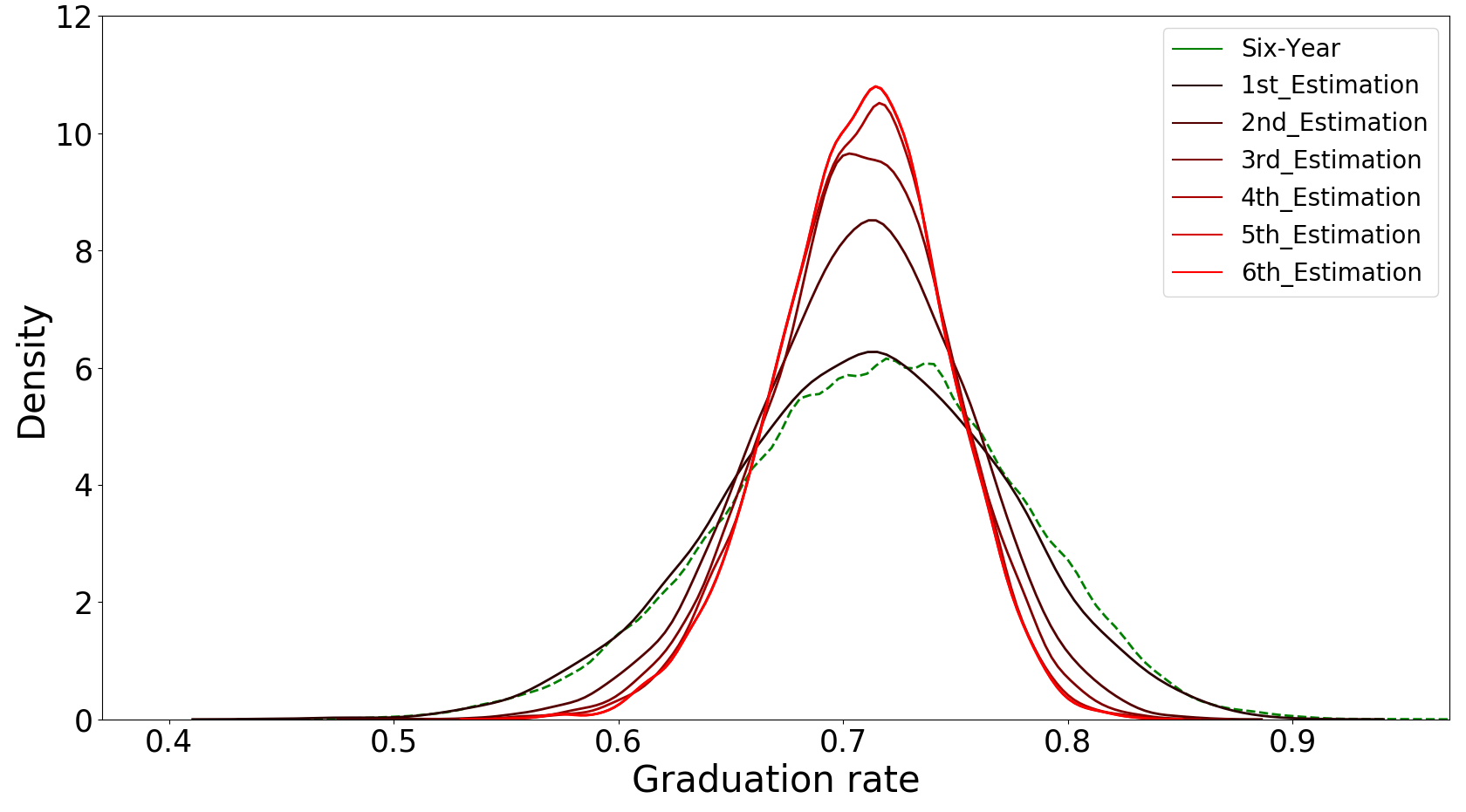}\hspace{5pt}

\caption{The probability density function (pdf) for estimations using the RUML-AMC with $n = 2$ (size = 50).} \label{fig:Density_Dynamic_approach}
\end{figure}

To decrease the sample standard deviation and estimation bias simultaneously, we apply the regularly updating multi-level absorbing Markov chain method with $n$ = 2.  Based on the resulting models, we then estimate graduation rates for cohorts with different sample sizes. Probability density functions of the first to sixth estimations for cohort with $N$ = 50 are shown in \figref{fig:Density_Dynamic_approach}. Each successive estimation is based on $1,~2,\hdots,~6$ incoming classes of students used to create the transition matrix for the  RUML-AMC. As~we see in the figure, the~approach has a strong performance in terms of limiting the estimation variation and estimation bias. \tabref{tab:std_comparison_MLAMC_different_complexities} provides the sample standard deviation of the estimates for the RUML-AMC with a different number of levels as well. The~results in \tabref{tab:std_comparison_MLAMC_different_complexities} demonstrate that when adding levels to an AMC to reduce the estimation bias (i.e., applying the technique from \secref{subsec:Bias}), the~byproduct is that the resulting model increases estimation variation; later in \secref{sec:Discussion}, we discuss the trade-off between estimation bias and estimation~variance.  

\begin{figure}[h!]
\centering
\includegraphics[width = 0.75\textwidth]{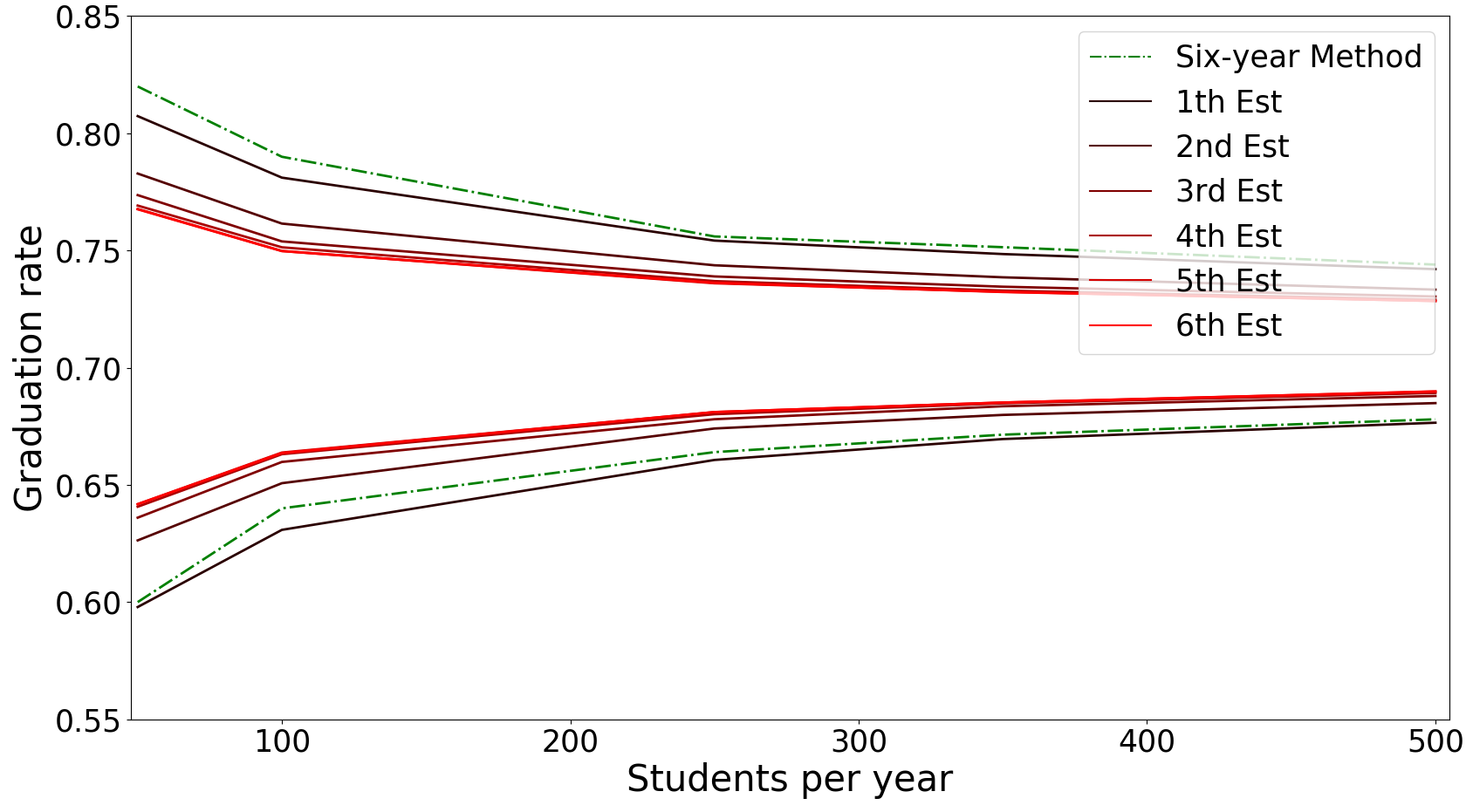}\hspace{5pt}
\caption{The 5--95\% inter-quartiles for graduation rates of cohorts with different sizes obtained by the RUML-AMC with $n$ = 2 and the six-year graduation rate~method.} \label{fig:dynamic_MC_Percentiles}
\end{figure}

\begin{table}[h!]
\caption{Standard deviation of the estimated graduation rate from years 1 to 6 for a cohort with $N$ = 50 for the RUML-AMC with different levels of~complexity.}
\label{tab:std_comparison_MLAMC_different_complexities}
\centering
\begin{tabular}{|c|c|c|c|c|c|c|}
\hline
 \textbf{Estimation Number} & \textbf{One Level} & \textbf{Two Levels} & \textbf{Three Levels}	\\ 
\hline
1 & 6.23\% & 6.52\% & 7.02\%   \\ \hline
2 & 4.47\% & 4.70\% & 5.53\%    \\ \hline
3 & 3.93\% & 4.18\% & 5.46\%    \\ \hline
4 & 3.61\% & 3.91\% & 4.80\%    \\ \hline
5 & 3.48\% & 3.77\% & 4.68\%    \\ \hline
6 & 3.39\% & 3.71\% & 4.39\%    \\ \hline
\end{tabular}
\end{table}
\unskip
\figref{fig:dynamic_MC_Percentiles} compares the 5--95\% inter-quartile intervals of the six estimations obtained by the RUML-AMC alongside the six-year graduation approach. As~illustrated in the figure, for~a fixed number of students added per year, the~gap between the 5--95\% inter-quartiles is reduced from the first to the sixth estimation. In addition, by~increasing the number of students added per year, the~estimations for the transition probabilities become increasingly accurate, along with the final graduation estimate. Based on these results, we can assert that our proposed approach has better performance than the SYGR method and standard AMC method when estimating graduation rates for cohorts with small sample~sizes.

Use real-world data, we can validate our proposed model (RUML-AMC) and compare results with the SYGR and AMC. To~do that, we choose the two colleges with the highest numbers of registrations for Fall 2008 at the University of Central Florida: the College of Science and the College of Engineering and Computer Science. We also consider students from all other colleges in another group and indicate it as {"Other colleges"}. We randomly select 500 students from each group who enrolled in Fall 2008 and compute the true value of the colleges' graduation rate with the SYGR method. To~evaluate the performance of the SYGR and AMC methods in terms of the standard deviation of estimated graduation rate, we~split the 500 students into ten cohorts of 50 students each and compute the graduation rate for each cohort using the SYGR and AMC. We also compute the graduation rate with our proposed RUML-AMC model with two sub-states ($n = 2$).  Since in this method, we regularly update the transition matrix over the course of six years, we need to consider the new students who join the program each year as well. Therefore, we add 50 new students to the pool of existing students every year from Fall 2008 to Fall 2013 and update the transition matrix. \tabref{tab:std_comparison_College_Science_ENG} provides the sample standard deviation and mean of the estimated graduation rates computed by Equations \eqref{eq:sample_std} and  \eqref{eq:sample_mean} for the 10 cohorts with $N = 50$ using the SYGR, AMC, and~RUML-AMC (forthe RUML-AMC, the~standard deviation of the sixth estimation is reported) with $n = 2$. As~the table shows, for~all cases, the~RUML-AMC has the lowest standard deviation of the estimated graduation rate compared to the SYGR and AMC methods. In addition, the~estimation bias has been decreased with the RUML-AMC with $n = 2$ compared to the AMC method.  Accordingly, we have demonstrated using real-world data that the proposed RUML-AMC method not only provides an accurate estimate of the graduation rate, but it also outperforms the SYGR and AMC methods.
\begin{equation} \label{eq:sample_std}
 s = \sqrt{\frac{1}{10-1}\sum_{i = 1}^{10}(x_i-\bar{x})^2}
\end{equation}
\begin{equation} \label{eq:sample_mean}
\bar{x} = \frac{1}{10}\sum_{i = 1}^{10} x_i
\end{equation}

\unskip
\begin{table}[h!]

\caption{Standard deviation of estimated graduation rates for 10 cohorts with $N$ = 50 using the SYGR, AMC, and~RUML-AMC with $n = 2$.}
\centering
\begin{tabular}{c@{\qquad}ccc@{\qquad}ccc}

\hline
\multirow{2}{*}{{\textbf{College}}} & \multicolumn{2}{c}{\textbf{SYGR}} &
\multicolumn{2}{c}{\textbf{AMC}}&
\multicolumn{2}{c}{\textbf{RUML-AMC}} \\

& \textbf{Std} & \textbf{Mean} & \textbf{Std} & \textbf{Mean} & \textbf{Std} & \textbf{Mean} \\
\hline
College of Science &  6.14\% & 71.2\% & 5.29\% & 69.6\% & 2.99\% & 71.0\% \\\hline
College of Engineering and CS& 5.48\% & 66.2\% & 5.34\% & 60.8\% & 3.48\% & 65.6\% \\\hline
Other Colleges & 5.71\% & 72.4\% & 5.21\% & 68.3\% & 3.12\% & 72.1\% \\
\hline
\end{tabular}\label{tab:std_comparison_College_Science_ENG}

\end{table}
\unskip
\section{Discussion}\label{sec:Discussion}
In the previous section, we observed that different levels of model complexity affect the accuracy of the results in terms of bias and variance. In~this section, we discuss the bias--variance trade-off and its effect on the estimation error. The~total estimation error for each method consists of two parts: bias error and variance error. Bias error is defined as the difference between model estimation and the true target value. Variance error tells us how the model estimations spread around the predicted mean. Typically, models with greater complexity have a higher variance error and a lower bias error. Therefore, it is critical to clarify our priority between minimizing the bias and variance when selecting a specific~model.

\begin{figure}[h!]
\centering
\includegraphics[width = 0.70\textwidth]{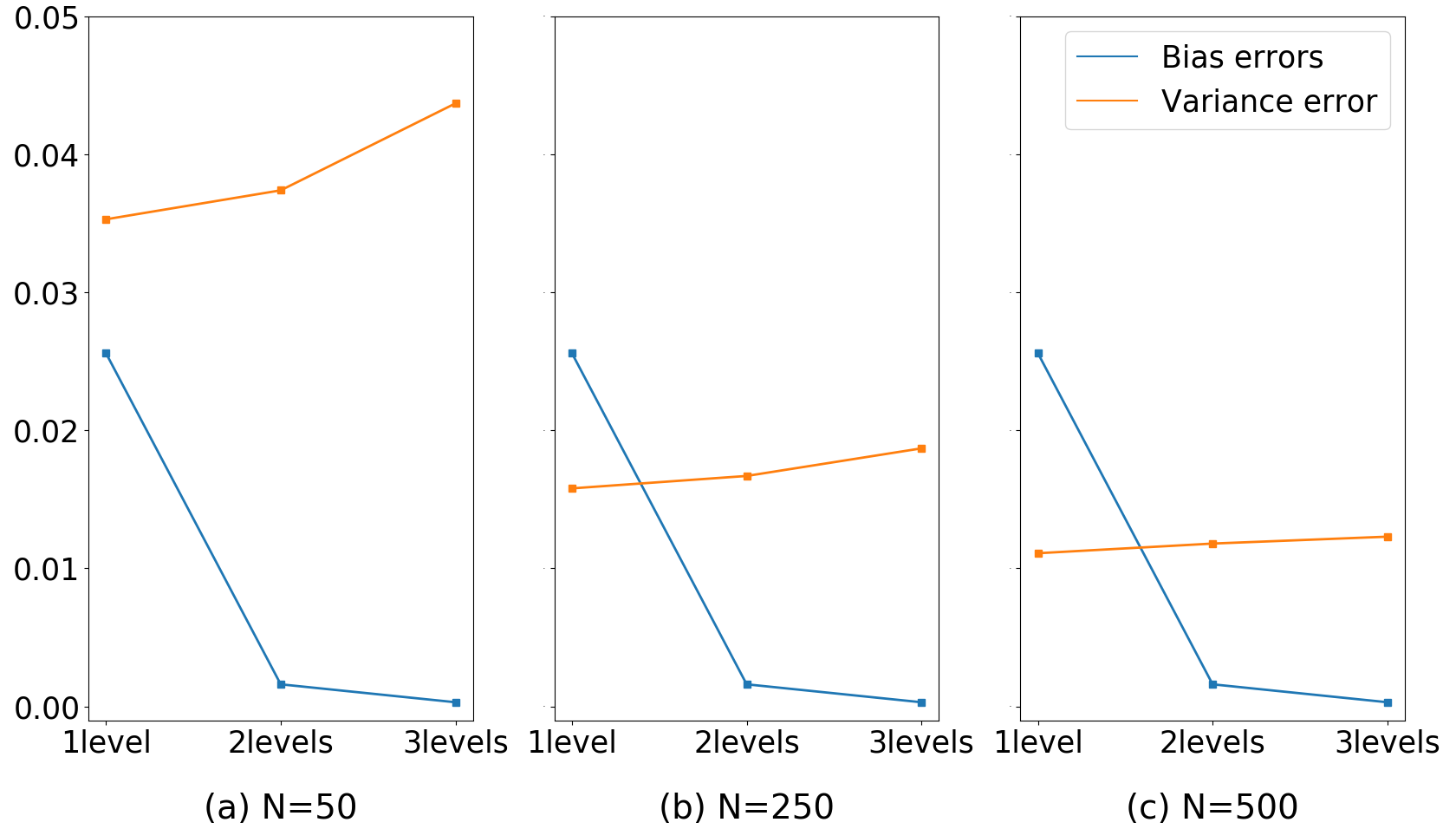}\hspace{5pt}
\caption{Bias, standard deviation, and~total error for models with different levels of complexity and different sample~sizes.} \label{fig:Bias_variance}
\end{figure}

\tabref{tab:std_comparison_MLAMC_different_complexities} compares the standard deviation for three different models with different levels of complexity. As~we see in the table, the~more complex the model (number of levels for each state), the~larger the variance error (as reported by the sample standard deviation) for the model. \figref{fig:Bias_variance} depicts the bias, standard deviation, and~total errors (sum of bias and variance errors) for the different models with different sample sizes. As~the figure suggests, for~the cohorts with sample sizes of 50~and 250, the~model with two levels has the lowest total error among the other models, and for the cohort with the sample size of 500, the~model with three levels has the best performance. For~the models discussed in this paper, we considered a consistent number of levels for all transient states. However, based on the context, this approach can be customized to include different numbers of levels for different states (i.e., Freshman, Sophomore, Junior, Senior).

Finally, it is worth discussing additional applications of the method presented here.  For~example, researchers can apply Equation~(\ref{eq:B_Mat}) to compute the absorption probability matrix B, whose values indicate the percentage of students who start at university and ultimately graduate, regardless of the number of years they are registered at the university.  This measure of the ultimate graduation rate allows universities to take a more holistic view of academic outcomes instead of focusing strictly on the six-year graduation rate. We apply this equation to the RUML-AMC model using the Fall 2008 incoming class.
The~estimated ultimate graduation rate is 74.41\%, which is similar to the percentage of students who earn a degree within 8 years of enrolling at the university, 74.47\% (since our data set covers eight years of data from 2008 to 2016, this percentage reports the eight-year graduation rate).  Similarly, researchers can apply this method to compute the graduation rate for transfer students as well, since this approach considers students that join the university with any academic level. The~computed two-year, three-year, four-year, five-year, six-year, and~overall graduation rates using the RUML-AMC with $n = 2$ for transfer students who joined the University of Central Florida in Fall 2008 are shown in \tabref{tab:Grate_for_Tr_students}. In~the table, the~column labeled "{retention rate}" provides the probabilities with which students will remain in school the next year. For~example, the~retention rate in the first row (second year) shows that 47.00\% of the transfer students will enroll at UCF in the third year (31.74\% graduate and 21.26\% halt by the end of the second year). In addition,~only~6.54\%~of the transfer students have enrolled for more than four years at UCF (the retention rate for the third row). Finally, the~overall graduation rate for transfer students based on Equation~(\ref{eq:B_Mat}) is 74.30\%, which means that 25.70\% of transfer students leave UCF without earning a degree regardless of the number of years they are enrolled at UCF; these results are consistent with the university's reported statistics~\cite{kruckemyer_2018}.  Perhaps the most beneficial potential application of the RUML-AMC is for estimating the graduation rates of students that enroll part-time, which, according to national statistics, is a large fraction of American college students~\cite{center2017even}, as~71\% of students alternate or fully complete their degrees using part-time semesters.  Without~longitudinal studies, it is challenging to measure the impact of policy decisions (e.g., financial aid, curriculum) on the long-term academic performance measures of graduation for part-time students.  The~RUML-AMC method proposed here provides a means to estimate graduation rates, even when data availability is~limited.

\begin{table}[h!]
\caption{The computed two-year, three-year, four-year, five-year, six-year, and~eventual graduation rates using the RUML-AMC with $n = 2$ for transfer students who joined the University of Central Florida in Fall~2008.}
\centering
\begin{tabular}{|c|c|c|c|}
\hline
 \textbf{Graduation Rate Method} & \textbf{
 Graduation Rate} & \textbf{Halt Rate} & \textbf{Retention Rate}	\\
\hline

Two-year & 31.74\% & 21.26\% & 47.00\%  \\ \hline
Three-year & 56.43\% & 24.07\%& 19.50\%   \\ \hline
Four-year & 68.43\%& 25.03\%& 6.54\%     \\ \hline
Five-year & 72.59\% & 25.34\%& 2.07\%    \\ \hline
Six-year & 73.92\% & 25.44\%& 0.64\%    \\ \hline
Overall & 74.30\% & 25.70\%& 0.00\%     \\ \hline

\end{tabular}\label{tab:Grate_for_Tr_students}

\end{table}
\unskip
\section{Conclusions}\label{sec:Conclusion}
This paper proposes using a regularly updating multi-level absorbing Markov chain method as an alternative to the six-year graduation rate method for computing students' graduation rate, especially when the sample population is small. In~the proposed approach, we make use of multiple levels for each transient state while updating the transition matrix year by year based on the existing and joining pool of students and their academic performances. With~these adjustments, we still maintain transition states of the Markov chain corresponding to students' academic level, and~the absorbing states are graduation and halting. Our sensitivity analysis shows that the estimated graduation rates obtained by the regularly updating multi-level absorbing Markov chain model give a more robust measure of graduation rate, even for a small population. For~a cohort with $N$  = 50 students, our proposed approach with two levels for each state ($n$ = 2) almost eliminates the bias error and reduces estimation variation by more than 40\% compared to the six-year graduation rate~method.

While the regularly updating multi-level Markov chain approach requires the inclusion of  data of students that are not in the same year as the initial entering class, we find this approach more appropriate than the standard SYGR. As~mentioned previously, the~SYGR is arguably a stale statistic. Assuming that graduation rates remain static through multi-year periods, then our proposed method is an improvement, as it can capture changes in graduation rates should there be significant shifts in the degree program---additionally, standard statistical hypothesis testing techniques can be applied to determine if the underlying transition matrix has evolved with~time.

\bibliographystyle{abbrv}
\bibliography{ref}

\begin{thebibliography}{10}

\bibitem{abbott2002influence}
M.~L. Abbott, J.~Joireman, and H.~R. Stroh.
\newblock The influence of district size, school size and socioeconomic status
  on student achievement in washington: A replication study using hierarchical
  linear modeling. technical report.
\newblock 2002.

\bibitem{aghajari2020decomposition}
Z.~Aghajari, D.~S. Unal, M.~E. Unal, L.~G{\'o}mez, and E.~Walker.
\newblock Decomposition of response time to give better prediction of
  children's reading comprehension.
\newblock {\em International Educational Data Mining Society}, 2020.

\bibitem{al2007application}
S.~A. Al-Awadhi and M.~Konsowa.
\newblock An application of absorbing markov analysis to the student flow in an
  academic institution.
\newblock {\em Kuwait J. Sci. Eng}, 34(2A):77--89, 2007.

\bibitem{bairagi2017stochastic}
A.~Bairagi and S.~C. Kakaty.
\newblock A stochastic process approach to analyse students' performance in
  higher education institutions.
\newblock {\em International Journal of Statistics and Systems},
  12(2):323--342, 2017.

\bibitem{boujelbenedata}
M.~Boujelbene and M.~K. Damak.
\newblock A data science approach to flagging non-retention in engineering
  enrollment data.

\bibitem{boumi2019application}
S.~Boumi and A.~Vela.
\newblock Application of hidden markov models to quantify the impact of
  enrollment patterns on student performance.
\newblock {\em International Educational Data Mining Society}, 2019.

\bibitem{boumi2020quantifying}
S.~Boumi, A.~Vela, and J.~Chini.
\newblock Quantifying the relationship between student enrollment patterns and
  student performance.
\newblock {\em arXiv preprint arXiv:2003.10874}, 2020.

\bibitem{brezavvsvcek2017markov}
A.~Brezav{\v{s}}{\v{c}}ek, M.~P. Bach, and A.~Baggia.
\newblock Markov analysis of students' performance and academic progress in
  higher education.
\newblock {\em Organizacija}, 50(2):83--95, 2017.

\bibitem{center2017even}
{Center for Community College Student Engagement}.
\newblock Even one semester: Full-time enrollment and student success.
\newblock Technical report, The University of Texas at Austin, College of
  Education, 2017.

\bibitem{ebrahiminejad2019pathways}
H.~EbrahimiNejad, H.~A. Al~Yagoub, G.~D. Ricco, M.~W. Ohland, and L.~Zahedi.
\newblock Pathways and outcomes of rural students in engineering.
\newblock In {\em 2019 IEEE Frontiers in Education Conference (FIE)}, pages
  1--6. IEEE, 2019.

\bibitem{eledum2019undergraduate}
H.~Eledum and E.~I.~M. Idriss.
\newblock An undergraduate student flow model: Semester system in university of
  tabuk (ksa).
\newblock 2019.

\bibitem{grace2017three}
K.~Grace-Martin.
\newblock Three issues in sample size estimates for multilevel models.
\newblock {\em The analysis factor: Making statistics make sense}, 2017.

\bibitem{hadad2020source}
S.~Hadad, M.~Dinu, R.~Bumbac, M.-C. Iorgulescu, and R.~Cantaragiu.
\newblock Source of knowledge dynamics—transition from high school to
  university.
\newblock {\em Entropy}, 22(9):918, 2020.

\bibitem{hagedorn2005define}
L.~S. Hagedorn.
\newblock How to define retention.
\newblock {\em College student retention formula for student success}, pages
  90--105, 2005.

\bibitem{heinrich2005parameter}
G.~Heinrich.
\newblock Parameter estimation for text analysis.
\newblock Technical report, Technical report, 2005.

\bibitem{kruckemyer_2018}
G.~Kruckemyer.
\newblock Ucf recognized for programs benefiting transfer students: University
  of central florida news, Nov 2018.

\bibitem{ledwith2019ethics}
M.~C. Ledwith, R.~A. Jackson, A.~M. Reboulet, and T.~P. Talafuse.
\newblock Ethics and education: A markov chain assessment of civilian education
  in air force materiel command.
\newblock {\em International Journal of Responsible Leadership and Ethical
  Decision-Making (IJRLEDM)}, 1(1):25--37, 2019.

\bibitem{maas2005sufficient}
C.~J. Maas and J.~J. Hox.
\newblock Sufficient sample sizes for multilevel modeling.
\newblock {\em Methodology}, 1(3):86--92, 2005.

\bibitem{mirzaei2019modeling}
M.~Mirzaei and S.~Sahebi.
\newblock Modeling students’ behavior using sequential patterns to predict
  their performance.
\newblock In {\em International Conference on Artificial Intelligence in
  Education}, pages 350--353. Springer, 2019.

\bibitem{mirzaei2020detecting}
M.~Mirzaei, S.~Sahebi, and P.~Brusilovsky.
\newblock Detecting trait versus performance student behavioral patterns using
  discriminative non-negative matrix factorization.
\newblock In {\em The Thirty-Third International Flairs Conference}, 2020.

\bibitem{nicholls2009use}
M.~Nicholls.
\newblock The use of markov models as an aid to the evaluation, planning and
  benchmarking of doctoral programs.
\newblock {\em Journal of the Operational Research Society}, 60(9):1183--1190,
  2009.

\bibitem{nicholls2007assessing}
M.~G. Nicholls.
\newblock Assessing the progress and the underlying nature of the flows of
  doctoral and master degree candidates using absorbing markov chains.
\newblock {\em Higher Education}, 53(6):769--790, 2007.

\bibitem{polyzou2019scholars}
A.~Polyzou, N.~Athanasios, and G.~Karypis.
\newblock Scholars walk: A markov chain framework for course recommendation.
\newblock In {\em Proceedings of the 12th International Conference on
  Educational Data Mining}, pages 396--401, 2019.

\bibitem{porter2019women}
A.~M. Porter and R.~Ivie.
\newblock Women in physics and astronomy, 2019. report.
\newblock {\em AIP Statistical Research Center}, 2019.

\bibitem{ross2014introduction}
S.~M. Ross.
\newblock {\em Introduction to probability models}.
\newblock Academic press, 2014.

\bibitem{saa2016educational}
A.~A. Saa.
\newblock Educational data mining \& students’ performance prediction.
\newblock {\em International Journal of Advanced Computer Science and
  Applications}, 7(5):212--220, 2016.

\bibitem{shah1999undergraduate}
C.~Shah and G.~Burke.
\newblock An undergraduate student flow model: Australian higher education.
\newblock {\em Higher Education}, 37(4):359--375, 1999.

\bibitem{WinNT}
K.~Too.
\newblock Largest universities in the united states by enrollment.
\newblock Technical report, 2017.

\bibitem{tugend2018colleges}
A.~Tugend.
\newblock Colleges and universities woo once-overlooked transfer students.
\newblock {\em New York Times}, 2018.

\bibitem{wuhib2014so}
F.~W. Wuhib and S.~Dotger.
\newblock Why so few women in stem: The role of social coping.
\newblock In {\em 2014 IEEE Integrated STEM Education Conference}, pages 1--7.
  IEEE, 2014.

\bibitem{zahedi2020multi}
L.~Zahedi, H.~Ebrahiminejad, M.~S. Ross, M.~Ohland, and S.~Lunn.
\newblock Multi-institution study of student demographics and stickiness of
  comput-ing majors in the usa.
\newblock {\em Collaborative Network for Engineering and Computing Diversity
  (CoNECD)}, 2020.

\bibitem{asee_peer_34921}
L.~Zahedi, S.~J. Lunn, S.~Pouyanfar, M.~S. Ross, and M.~W. Ohland.
\newblock Leveraging machine-learning techniques to analyze computing
  persistence in undergraduate programs.
\newblock In {\em 2020 ASEE Virtual Annual Conference Content Access}, number
  10.18260/1-2--34921, Virtual On line, June 2020. ASEE Conferences.
\newblock https://peer.asee.org/34921.

\end{thebibliography}

\end{document}